\documentclass[fleqn,usenatbib]{mnras}

\usepackage[T1]{fontenc}
\usepackage{ae,aecompl}
\usepackage{xcolor}
\usepackage{comment}

\usepackage{graphicx}	
\usepackage{amsmath}	
\usepackage{amssymb}	
\usepackage{pdflscape}	
\usepackage[english]{babel}
\pdfoutput=1

\title[Planet formation and the dust budget crisis]{Can planet formation resolve the dust budget crisis in high redshift galaxies?}

\author[D. H. Forgan et al. ]
{D. H. Forgan$^{1,2}$\thanks{Contact e-mail: \href{mailto:dhf3@st-andrews.ac.uk}{dhf3@st-andrews.ac.uk}},
K. Rowlands$^{1,3}$, H. L. Gomez$^{4}$, E. L. Gomez$^{5,4}$, S. P. Schofield$^{4}$, \and L. Dunne$^{4,6}$, S. Maddox$^{4,6}$
\vspace{0.2cm} \\
$^{1}$SUPA, School of Physics \& Astronomy, University of St Andrews, North Haugh, St Andrews, Scotland, KY16 9SS, UK\\
$^{2}$St Andrews Centre for Exoplanet Science, University of St Andrews, UK \\
$^{3}$Department of Physics \& Astronomy, Johns Hopkins University, Bloomberg Center, 3400 N. Charles St., Baltimore, MD 21218, USA \\
$^{4}$School of Physics \&\ Astronomy, Cardiff University, Queens Buildings, The Parade, Cardiff, CF24 3AA, UK \\
$^{5}$Las Cumbres Observatory, Suite 102, 6740 Cortona Dr, Goleta, CA 93117, USA \\
$^{6}$SUPA, Institute for Astronomy, University of Edinburgh, Royal Observatory, Blackford Hill, Edinburgh, EH9 3HJ, UK
}
\date{Accepted XXX. Received YYY; in original form ZZZ}

\pubyear{2017}

\begin{document}
\label{firstpage}
\pagerange{\pageref{firstpage}--\pageref{lastpage}}
\maketitle

\begin{abstract}
The process of planet formation offers a rich source of dust production via grain growth in protostellar discs, and via grinding of larger bodies in debris disc systems. Chemical evolution models, designed to follow the build up of metals and dust in galaxies, do not currently account for planet formation. We consider the possibility that the apparent under-prediction of dust mass in high redshift galaxies by chemical evolution models could be in part, due to these models neglecting this process, specifically due to their assumption that a large fraction of the dust mass is removed from the interstellar medium during star formation (so-called astration). By adding a planet formation phase into galaxy chemical evolution, we demonstrate that the dust budget crisis can be partially ameliorated by a factor of 1.3-1.5 only if a) circumstellar discs prevent a large fraction of the dust mass entering the star during its birth, and b) that dust mass is preferentially liberated via jets, winds and outflows rather than accreted into planetary-mass bodies.  
\end{abstract}

\begin{keywords}
galaxies: high-redshift, evolution, abundances  --- planets and satellites: formation --- methods: numerical
\end{keywords}



\section{Introduction}
\label{sec:introduction}

The dust budget crisis in galaxies was first shown in \citet{Morgan2003}, where the amount of dust from stellar winds of low-intermediate mass stars (believed to be the dominant source of stardust in galaxies, \citealt{Whittet2003}), failed to explain the quantities observed in high redshift galaxies ($z \sim 5$). Since then, huge dusty reservoirs discovered at $z\sim 6-7$ with the \emph{Herschel Space Observatory} \citep{Riechers13} and ALMA \citep{Watson2015, Willott15, Knudsen16} place even more stringent constraints on dust sources in galaxies, requiring significant dust production on timescales of only a few hundred million years \citep{Gall2011, Mattsson2015, Michalowski2010, Mancini2015}.

Moving away from individual galaxies, \citet{Rowlands2014b} used a chemical evolution model to follow the build up of gas and dust for the largest compiled sample of submillimetre galaxies at high redshifts ($1<z<5$). They found that even if dust destruction by supernova shock waves in the interstellar medium (ISM) is set to zero, winds from low-intermediate mass stars produce a shortfall in  dust mass by a factor of $\sim$240, for their sample of 26 galaxies. This shortfall cannot be explained by uncertainties in the dust mass loss rate in stellar winds, or by changing the initial mass function as originally suggested in \citet{Valiante2009} (this result was also found in \citealt{Gall2011} and \citealt{Mattsson2011} on smaller galaxy samples). Perhaps most surprisingly, this result was shown to be robust to changes in the star formation history and changes in the physical properties of the dust ($\kappa$). Adding supernovae as an extremely efficient dust source (implying no dust destruction in the shocks) allows us to reduce the shortfall to a factor of 6 on average, compared to the factor of 240 with dust from AGB stars only.

The possibility that the dust budget crisis is an artefact due to uncertainties in the observations e.g. gas and dust masses, initial mass functions etc. was raised in \citet{Mattsson2011} and \citet{Jones2011}.  However, in the post-\emph{Herschel} and ALMA era, it is more difficult to explain this shortfall due to measurement uncertainties given well constrained far-infrared emission for larger samples, and detailed comparison studies of spectral energy distribution fitting techniques showing these lead to only a factor of a few in dust mass uncertainty \citep{Amblard2014,Mattsson2015a}. Indeed, this issue is also not just seen in extreme, rare systems at high redshifts where observational uncertainties are potentially large, the dust `crisis' is also observed in our own neighbourhood. \citet{Dunne2011} and \citet{Rowlands2012} were unable to explain the dust masses of low redshift galaxies detected by \emph{Herschel} using chemical evolution models with Milky Way-like metal or dust yields or star formation efficiencies. When carrying out a census of dust sources in the Large and Small Magellanic Clouds, \citet{Matsuura2009} and others \citep{Boyer2009,Srinivasan2010,Boyer2012} showed that stellar winds from low-intermediate mass stars are responsible for less than 10\,per\,cent of the total dust mass. 

Two potential dust sources have been put forward to try and solve this crisis. First, recent far-infrared and submillimetre observations have discovered large quantities of dust within supernovae ejecta in both young remnants \citep{Matsuura2011,Matsuura2015} and historical sources \citep{Dunne2003,Rho2008,Gomez2012,Barlow2010,DeLooze2017}. Although it is unclear how much of this dust survives in the long term due to the expected sputtering and shattering of grains via supernovae shocks \citep{Bianchi2007,Bocchio2016}. If roughly half of the grains survive their journey into the ISM, this would alleviate the crisis in nearby sources. Alternatively, supernovae and stellar winds may be important for providing dust seeds, which are then grown in cool, dense regions of the interstellar medium \citep{Draine2009}. Indeed many recent studies trying to explain the observed dust masses and dust-to-metallicity gradients in nearby galaxies (e.g. Andromeda, \citealt{Mattsson2014}) and the earliest galaxies \citep{Wang17} suggests that grain growth (via accretion) would need to be responsible for $\sim$90\,per\,cent of dust mass  \citep{Dunne2011,Gall2011,Asano2013,Mattsson2012,Mattsson2014,Zhukovska2014,Michalowski2015}.  \citet{Rowlands2014b} showed that a combination of dust injected by supernovae, with significant freshly formed dust from interstellar grain growth can explain the missing factor of $\sim 240$ dust in their sample. 
The above discussion suggests that a number of potential different factors could be combined in order to help resolve the dust budget crisis, including pinning down uncertainties in the injected dust masses from stars and the timescales for grain growth in the ISM, the dust absorption properties, the initial mass function and uncertainties in the chemical evolution models used to predict the dust budgets of galaxies. However, one potential question that has been ignored to date in chemical evolution models is whether the planet formation process could reduce the amount of dust presumed to be locked up in stars. These models could therefore be ignoring a potentially rich source of solids at a variety of sizes. More importantly, they do not account for important physical processes that are demonstrably efficient at grain growth.

The mechanisms that drive protostellar disc evolution and planet formation are multitudinous and interdependent \citep{Haworth2016}. However, we are sufficiently informed by extensive multiwavelength observations of young stellar objects \citep{Aurora2016} and even protoplanets in formation \citep[e.g.][]{Kraus2012} that we are able to give a reasonable account of the physics at play \citep[see][and references within]{Dutrey2014,Dunham2014}.

Protostellar discs are a natural consequence of the star formation process.  During the collapse of prestellar cores into star systems, the excess angular momentum in the system forces a significant fraction of the total mass into the disc.  The relatively high surface density of the disc, and initially low relative velocity, allows grains to collide and grow, sedimenting to the midplane \citep{Goldreich1973} and beginning the core accretion process of planet formation \citep{Pollack1996}. 

Grains can continue to grow to the ``pebble regime'', where they then become subject to strong gas drag forces which cause rapid inward migration toward the central star \citep{Weidenschilling1977}.  Bodies that are fortunate enough to grow beyond this barrier can efficiently accrete pebbles and become protoplanets \citep{Lambrechts2014,Bitsch2015}.  If the protoplanet is sufficiently massive, it can accrete a gaseous envelope and become a gas giant.

During this process, the inner disc is continually being eroded by photoevaporative winds \citep[see][and references within]{Alexander2014}. Also, significant quantities of disc dust and gas are launched back into the ISM through jets and outflows \citep[][and references within]{Frank2014}.  Once the photoevaporative flow has penetrated deep enough into the disc, the remaining gaseous material is rapidly swept away, with some of the dust entrained in this wind.  Gas giant planet formation is halted at this stage, although terrestrial planets can continue to grow via dust and planetesimal accretion, and giant impacts \citep[e.g.][]{Quintana2016}.

The efficiency of this mechanism is evidenced by recent exoplanet detection surveys.  The combined constraints from radial velocity surveys \citep{Mayor2011}, transit surveys by the \emph{Kepler} space telescope \citep{Howard2012} and microlensing \citep{Cassan2012} suggest virtually every star has at least one planet, and by implication substantial amounts of debris in a range of grain sizes from microns to kilometres \citep[see also][]{Winn2015}.

This high efficiency partially exacerbates the dust budget crisis, as it locks a significant fraction of the dust mass into grains far bigger than observations are able to probe. However, the post-gas phase of the system can provide another source of dust. These debris discs continue to produce grains in the observable window by the grinding of kilometre-sized bodies as they collide \citep[see][for a review]{Matthews2014}. These systems are a continuing dust production source, essentially from a few Myr after the star forms up to several Gyr afterwards.

In this work, we will implement a simple toy model of planet formation into the chemical evolution code of \citet{Morgan2003, Rowlands2014b}, to understand broadly how planet formation affects the dust budget crisis. In section \ref{sec:method}, we describe our modifications to the code; in section \ref{sec:results} we simulate the dust evolution of several galaxies under various assumptions about the efficiency of planet formation and discuss the prospects for resolving (or worsening) the dust budget crisis; and in section \ref{sec:conclusions} we summarise the work.

\section{Method}
\label{sec:method}

From \citet{Rowlands2014b}, the chemical evolution model tracks the build-up of heavy elements over time produced by stars where some fraction of metals will condense into dust. Given an input star formation history (SFH), gas is converted into stars over time, assuming an initial mass function (IMF). The total mass of the system is given by

\begin{equation}
M_{\rm total}=M_g+M_*,
\label{eq:all}
\end{equation}

\noindent where {\bf $M_g$} is the gas mass and {\bf $M_*$} is the stellar mass. The gas mass changes with time as described in Eq.~\ref{ginc}, as gas is depleted by the SFR, $\psi(t)$, and returned to the ISM as stars die, $e(t)$:

\begin{equation}
{\frac{dM_g}{dt}} = -\psi(t) + e(t) + I(t) - O(t).
\label{ginc}
\end{equation}

\noindent The first two terms in Eq.~\ref{ginc} on their own describe a closed box system, the third term describes gas inflow with rate $I$ and the fourth term describes outflow of gas with rate $O$. Inflows and outflow rates are (in the simplest form) parameterised as a fraction of the instantaneous SFR. Assuming that mass loss occurs suddenly at the end of stellar evolution, the ejected mass is $e(t)$.

The evolution of the mass of metals in the ISM ({\bf$M_Z$}) is described by 

\begin{equation}
\frac{d(M_Z)}{dt} = -Z(t)\psi(t) + e_z(t) + Z_{I}I(t) - Z_{O}O(t)+ M_{Z,i},
\label{eq:metals}
\end{equation}

\noindent where $Z$ is defined as the fraction of heavy elements {\it by mass} in the gas phase ($M_Z/M_g$).  The first term of Eq.~\ref{eq:metals} describes the metals locked up in stars, and the second term describes the metals returned to the ISM via stellar mass loss, $e_z(t)$:

\begin{eqnarray}
e_z(t)&=& \int_{m_{\tau_m}}^{m_U}\bigl({\left[m-m_{R}(m)\right] Z(t-\tau_m)+mp_z}\bigr)  \nonumber \\ 
 & &\mbox{} \times \psi(t-\tau_m)\phi(m)dm
\label{eq:chemz}
\end{eqnarray}

\noindent where $mp_z$ is the yield of heavy elements from a star of initial mass $m$, remnant mass, $m_{R}(m)$ and metallicity $Z$. This process depends on the star formation rate $\psi$ at time $t-\tau_m$, where $\tau_m$ is the lifetime of a star with mass $m$, integrated over the IMF $\phi(m)$. The third term of Eq.~\ref{eq:metals} describes an inflow of gas with metallicity $Z_{I}$ and the fourth term of Eq.~\ref{eq:metals} describes an outflow of gas with metallicity $Z_{O}$. The final term $M_{Z,i}$ allows for pre-enrichment from Pop III stars, which is set to zero in this work.

The evolution of dust mass with time is given by:
 
\begin{multline} \label{eq:dustmass}
    \frac{dM_d}{dt} = \int^{m_U}_{m_{\tau_m}} \left(\left[m - m_R(m)\right]\right. \left.Z(t-\tau_m) \delta_\mathrm{lims} +m p_z \delta_\mathrm{dust} \right) \\
    \left.\times \psi(t-\tau_m) \phi(m)\right)dm -A_d(t) \\
    - \left(1-f_c\right)M_d\delta_\mathrm{dest}(t)  + f_c\left(1-\frac{M_d}{M_Z}\right)M_d\delta_\mathrm{grow}(t) +M_{d,i} + \\
    \left(\frac{M_d}{M_g}\right)_I I(t) - \left(\frac{M_d}{M_g}\right)_O O(t) 
\end{multline}

\noindent The first two terms of Eq.~\ref{eq:dustmass} correspond to the following scenario: metals are locked up in stars, and are then recycled into dust via stellar winds, alongside freshly synthesised heavy metals forged in low to intermediate mass stars (LIMS) and supernovae (SNe). Again, these processes depend on the star formation rate at time $t-\tau_m$ and the IMF.

The other terms in the above equation represent, in order, the astration of dust ($A_d$, i.e. $(M_d/M_g)\psi(t)$, the draw down of grains into stars forming at that instant). Dust destruction via SN shocks which occurs on timescales of $\delta_{\rm des}^{-1}$ and is an efficient destruction process for grains in the warm, diffuse phase of the ISM ($1-f_c$) where $f_c$ is the fraction of the ISM in the cold, molecular phase (set to 0.5 as default, \citealt{Inoue2003,Mancini2015}). The next two terms account for dust sources via grain growth (via accretion) in the ISM and dust produced by Pop III stars. The former process depends on the amount of metals available in the cold ISM (where growth occurs) that are not already locked up in dust grains $f_c(1-M_d/M_Z)$ with characteristic timescale $\delta_{\rm grow}^{-1}$.  The latter term is set to zero. The remaining terms account for dust mass gain/loss by gas inflow/outflow{\footnote{for more details on the parameters, see the full discussion in \citet{Rowlands2014b} and the changes since that work described in \citet{DeVis2017}.}. 

\subsection{Planet formation as a dust source and sink}
 
In chemical evolution models, the astration term is effectively set to 100\,per\,cent efficient at removing dust as that material forms stars, representing a significant dust sink in the above equation. Here, we argue that there may be modifications to this equation due to the planet formation process. In particular, the astration term will require significant modification. To do this, we propose a simplified toy model of planet formation which allows us to parametrise the quantity of dust that is produced and destroyed by protoplanetary discs during the birth of planetary systems. 
Note that we are including planet formation as separate metals and dust source terms in Eqs.~\ref{eq:metals} and \ref{eq:dustmass}. This has essentially the same effect as modifying the astration term} to account for any dust that is not wholly swallowed up in star formation, but is instead returned due to planet formation processes (e.g. formation and ejection of disc material).

Consider an individual star system forming from a giant molecular cloud. Thanks to angular momentum conservation, the collapsing prestellar core forms a protostar-protostellar disc system.  Simulations of protostellar collapse indicate that the disc in fact forms first \citep{Bate2010,Tsukamoto2015}.  Dissociation of $H_2$ then allows the protostar to form at the centre.  In this early epoch, the disc and star are approximately equal in mass.  Much of the disc mass will be accreted by the star thanks to efficient angular momentum transport through the disc's gravitational instability \citep[e.g.][]{Laughlin1994,Forgan2011}.  The rest will either be ejected from the system through magnetocentrifugal winds, jets and outflows, or assembled into planets, satellites and other bodies with a range of sizes, from kilometres to microns.

Our model has the following parameters: $f_\mathrm{disc}$, the fraction of the disc material that is never accreted by the star; $f_\mathrm{debris}$, the further fraction of unaccreted disc material that remains in grain sizes amenable to observations; $f_\mathrm{wind}$, the fraction of disc material that is ejected and returned to the ISM; finally, $f_\mathrm{planet}$ describes the further fraction of disc material that assembles into bodies larger than the observation window.  We demand

\begin{equation}
f_\mathrm{planet} + f_\mathrm{debris} + f_\mathrm{wind} = 1
\end{equation}

\noindent We can reduce the number of free parameters by assuming a size distribution of grains in the disc:

\begin{equation}
P(s) \propto s^{-3.5}
\label{eq:size}
\end{equation}

\noindent This is a simple relation that holds reasonably well for debris disc systems \citep[see e.g.][and references within]{Dohnanyi1969,Matthews2014}.  We then use this distribution to compute the mass in grains between two grain sizes $[s_1,s_2]$:

\begin{equation}
M[s_1,s_2] \propto \int_{s_1}^{s_2} \frac{4}{3} \pi \rho_s s^3 P(s) ds = M_{tot} \frac{s^{1/2}_2 - s^{1/2}_1}{s^{1/2}_\mathrm{max} - s^{1/2}_\mathrm{min}}
\end{equation}

\noindent By computing the fraction of the total mass in grains that correspond to the observable window, we estimate that $f_\mathrm{debris}$ is approximately 10\% of the sum of $f_\mathrm{planet}$ and $f_\mathrm{debris}$.

We are therefore left with two free parameters: the fraction of material in the star formation process that does not enter the protostar, $f_\mathrm{disc}$, and the subsequent fraction that is ejected from the system during planet formation, $f_\mathrm{wind}$. Once these are specified, we can state

\begin{align*}
f_\mathrm{planet} & = 0.9(1-f_\mathrm{wind}) \\
f_\mathrm{debris} & = 0.1(1-f_\mathrm{wind})
\end{align*}

\noindent and the mass of dust in the protoplanetary disc is:

\begin{equation}
\left(\frac{dM_d}{dt}\right)_\mathrm{protoplanetarydisc} = f_\mathrm{disc} f_\mathrm{debris} \psi(t) \frac{M_d}{M_g}.
\end{equation}

\noindent Note that material locked up in planets is not visible at far-infrared--millimeter wavelengths, hence no contribution from $f_\mathrm{planet}$. We should also add an injection term for dust being removed from the protostellar disc:

\begin{equation}
\left(\frac{dM_d}{dt}\right)_\mathrm{disc~winds} = f_\mathrm{disc} f_\mathrm{wind} \psi(t) \frac{M_d}{M_g}.
\end{equation}

\noindent We can rewrite this as a composite term which incorporates all planet formation processes as:

\begin{equation}
\left(\frac{dM_d}{dt}\right)_{\mathrm{z, protoplanetarydisc}} = f_\mathrm{disc} (f_\mathrm{wind}+f_\mathrm{debris}) \psi(t) \frac{M_d}{M_g}
\label{eq:newastrationterm_dust}
\end{equation}

\noindent Similarly, the metals term is

\begin{equation}
\left(\frac{dM_z}{dt}\right)_{\mathrm{d, protoplanetarydisc}} = f_\mathrm{disc} (f_\mathrm{wind}+f_\mathrm{debris}) \psi(t) \frac{M_Z}{M_g}
\label{eq:newastrationterm_metals}
\end{equation}

In this work we adapt the code\footnote{The python code used is open source and is available on {\sc github}: \url{https://github.com/zemogle/chemevol/releases/tag/v_forgan2017}, release version v\_forgan2017.} from \citet{Morgan2003}, \citet{Rowlands2014b}, and more recently in \citet{DeVis2017} so that the mass of metals and dust in the protostar is separate from that in the protoplanetary disc. We achieve this by adding Eq.~\ref{eq:newastrationterm_dust} as a new source term in Eq.~\ref{eq:dustmass}, which has the effect of reducing the amount of astrated dust. Terms such as the mass of dust involved in grain growth, destruction, in inflows and outflows in Eq.~\ref{eq:dustmass} are unmodified because they depend on the dust mass in the ISM, which includes the dust in the protoplanetary disc which is instantaneously released into the ISM after star formation.

Note that the metal mass is also modified in a similar way by Eq.~\ref{eq:newastrationterm_metals}, which affects both Eqs.~\ref{eq:metals} to \ref{eq:dustmass}. We subtract the mass of metals in the star-disc system before star formation by putting it in the protoplanetary disc. Therefore we define a modified metallicity term

\begin{equation}
Z_\dag = Z - f_\mathrm{disc} (f_\mathrm{wind}+f_\mathrm{debris})Z
\label{eq:metallicity_mod}
\end{equation}

\noindent The metals now evolve as

\begin{equation}
\frac{d(M_Z)}{dt} = -Z_\dag(t)\psi(t) + e_{Z_\dag}(t) + Z_{I}I(t) - Z_{O}O(t)+ M_{Z,i},
\label{eq:metals_new}
\end{equation}

The modification means that the astrated mass is slightly less due to dust being in the protoplanetary disc, but fewer metals are ejected into the ISM at each timestep (from both freshly formed and recycled metals). Terms such as the mass of metals involved in inflows and outflows are unaffected, because the metals in the disc are released into the ISM instantaneously after star formation, so that mass is conserved.
We can recover the original dust and metals equations by setting $f_\mathrm{disc}=0$.

The evolution of dust mass with time is then:
 
\begin{multline} \label{eq:dustmass_new}
    \frac{dM_d}{dt} = \int^{m_U}_{m_{\tau_m}} \left(\left[m - m_R(m)\right]\right. \left.Z_\dag(t-\tau_m) \delta_\mathrm{lims} +m p_z \delta_\mathrm{dust} \right) \\
    \left.\times \psi(t-\tau_m) \phi(m)\right)dm -(1-f_\mathrm{disc} (f_\mathrm{wind}+f_\mathrm{debris})A_d \\
    - \left(1-f_c\right)M_d\delta_\mathrm{dest}(t)  + f_c\left(1-\frac{M_d}{M_Z}\right)M_d\delta_\mathrm{grow}(t) +M_{d,i} + \\
    \left(\frac{M_d}{M_g}\right)_I I(t) - \left(\frac{M_d}{M_g}\right)_O O(t)
\end{multline}

The gas astration term is not affected by planet formation, as the vast majority of disc gas mass is accreted by the star via the disc (i.e. the gas retained by giant planets or lost in winds is negligible). A substantial fraction of the stellar mass that constitutes the IMF is assembled by disc accretion, even at relatively large stellar masses \citep[cf][]{Johnston2015,Ilee2016}. Previously, astration terms did not account for the fact that planet formation processes significantly alter the gas-to-dust ratios in the material accreting onto the star. In our revised model, the IMF is relatively unchanged, but the accretion of dust and metals is altered by our knowledge that most stars possess planetary systems and/or debris. We then rerun the models using different values of $f_\mathrm{disc}$ and $f_\mathrm{wind}$ for three fiducial SFHs and for those galaxies in \citet{Rowlands2014a, Rowlands2014b}. 

Before we discuss the results of this work in section~\ref{sec:results}, we briefly highlight a different potential motivation for modifying the astration term in Eq.~\ref{eq:dustmass} following the simulations of \citet{Hopkins2016}. In their work, they found that the gas and dust can decouple in molecular clouds: at low densities, they predict that the dust-to-gas ratio is larger, and at high densities, they show that the dust follows the gas on average, but with significant fluctuations in this ratio. Since chemical evolution models assume that dust is astrated from the system at the current dust-to-gas ratio, this decoupling could potentially reduce or increase the amount of dust/metals lost due to astration. In this scenario, one would also need to modify the model grain growth and dust destruction timescales, as both terms would have account for an enhanced dust-dust interaction rate in the densest regions or where the largest grains are located (we note that the effect on the grain growth timescale might not be so drastic given the large grains are more affected). On a global, integrated level we account for this using the dense cloud fraction $f_c$, but we do not take this into account in the astration term.

As \citet{Hopkins2016} state that the fluctuations in the dust-to-gas ratio will smooth out on larger scales than the molecular clouds, where one will simply be tracing the mean dust-to-gas ratio, we do not modify the dust and metal astration terms further in this work.

\section{Results \& Discussion}
\label{sec:results}

\subsection{Dependence on the planet formation parameters}

We begin by running the chemical evolution model in a parameter sweep over $f_\mathrm{disc}$ and $f_\mathrm{wind}$, which in turn determine the other planet formation parameters $f_\mathrm{planet}$ and $f_\mathrm{debris}$.  In previous work where the above parameters are effectively zero, \citet{Rowlands2014b} found that dust from low-intermediate mass stars, supernovae and grain growth  (via accretion) is required in order to reproduce the observed dust masses of galaxies, with grain growth being the dominant dust producer. To see if planet formation processes can alleviate the dust budget crisis without grain growth, we consider dust produced by low-intermediate mass stars and supernovae in the chemical evolution model, and assume dust is not destroyed by supernova shocks.

For this initial sweep, we follow \citet{Rowlands2014b} and select three different  SFHs - a Milky Way-like SFH, a SFH with delayed star formation, and a burst SFH. For each SFH, we calculate the total dust mass produced, both in the presence and absence of planet formation (we label these as M$_\mathrm{dust,0}$ and M$_\mathrm{dust,p}$ respectively). We then calculate the relative increase in dust mass due to planet formation as

\begin{equation}
    R_{\rm dust} = \frac{\mathrm{M_{dust},p}}{\mathrm{M_{dust},0}}
\end{equation}

\begin{figure*}
\includegraphics[width=0.32\textwidth, clip=true, trim=4mm 4mm 4mm 3.5mm]{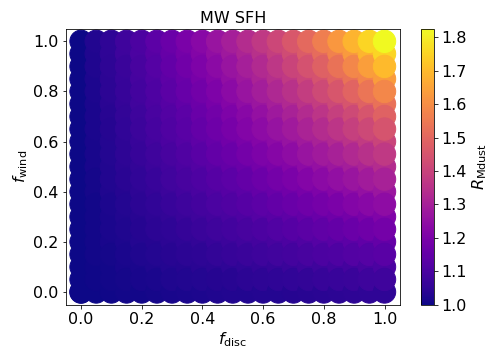}
\includegraphics[width=0.32\textwidth, clip=true, trim=4mm 4mm 4mm 3.5mm]{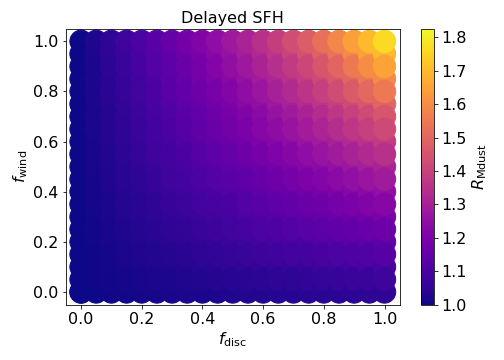}
\includegraphics[width=0.32\textwidth, clip=true, trim=4mm 4mm 4mm 3.5mm]{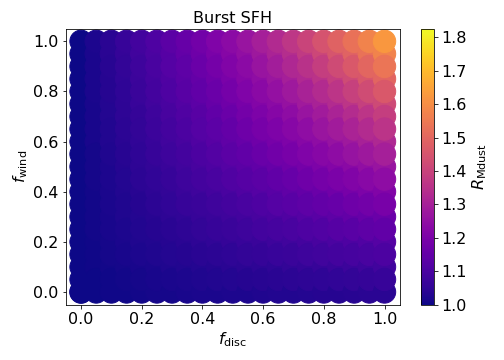}
\caption{The excess dust produced by planet formation compared to no planet formation (i.e. normal astration), for different values of $f_\mathrm{disc}$ and $f_\mathrm{wind}$, \emph{at time of maximum dust mass in the no planets case} (i.e. M$_\mathrm{dust},0 = $max, see the dashed line in Figure~\ref{fig:Extradust_planets_time}). Three different star-formation histories are shown: Milky Way (left), delayed (middle), burst (right). The models assume no dust destruction by supernova shocks.}
\label{fig:Extradust_planets}
\end{figure*}

\begin{figure*}
\includegraphics[width=0.32\textwidth, clip=true, trim=4mm 4mm 4mm 3.5mm]{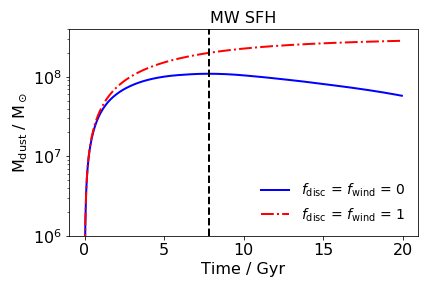}
\includegraphics[width=0.32\textwidth, clip=true, trim=4mm 4mm 4mm 3.5mm]{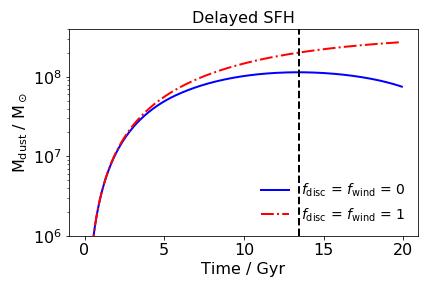}
\includegraphics[width=0.32\textwidth, clip=true, trim=4mm 4mm 4mm 3.5mm]{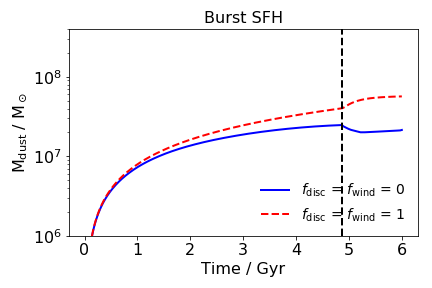}
\caption{Dust mass produced by maximal planet formation (red line, $f_\mathrm{disc}=1.0$ and $f_\mathrm{wind}=1.0$) compared to no planet formation (blue line), as a function of time. Three different star-formation histories are shown: Milky Way (left), delayed (middle), burst (right). The dashed lines indicate the time of maximum dust production in the absence of planet formation.  The models assume no dust destruction by supernova shocks.}
\label{fig:Extradust_planets_time}
\end{figure*}

\noindent Which is plotted as a function of $(f_\mathrm{disc},f_\mathrm{wind})$ in Figure~\ref{fig:Extradust_planets} for each SFH. Figure~\ref{fig:Extradust_planets_time} shows the time evolution of the dust mass for each SFH, given the maximal parameters $f_\mathrm{disc} = 1$, $f_\mathrm{wind} = 1.0$. We calculate $R_\mathrm{dust}$ at the time where M$_\mathrm{dust,0}$ reaches its maximum (indicated by the dashed lines in Figure~\ref{fig:Extradust_planets_time}), i.e. at 7.8, 13.5 and 4.9 Gyrs after the onset of star formation in the Milky Way, delayed and burst SFHs, respectively.

In all three cases, all values of $f_\mathrm{disc}$ and $f_\mathrm{wind}$ result in less than a factor of two increase in the dust mass. For the Milky Way SFH (left plot of Figure~\ref{fig:Extradust_planets}), values of $f_\mathrm{disc}$ and $f_\mathrm{wind}$ greater than 0.7 - i.e. a reduction of 70\% in the drawdown of dust into stars, with 70\% of the dust mass not drawn into the star being liberated through jets and outflows - results in more than a factor of 1.3 increase in the dust mass. A factor of 1.8 more dust is produced if none of the dust is accreted by the star, and none of that dust forms planets. This is not enough to make a significant difference to the dust budget crisis if dust is only produced by low-intermediate mass stars and supernovae. 

The increase in the amount of dust depends slightly on the SFH, with the delayed SFH (middle plot) showing less of an increase in the dust mass with planet formation compared to the Milky Way. The burst SFH (right plot) shows the smallest increase in the dust mass, with a factor of 1.7 more dust produced if none of the dust is accreted by the star, and none of that dust forms planets, which is clearly unphysical. Whether or not we run the model with grain growth makes a negligible difference to the amount of extra dust produced by planet formation. The increase in dust mass due to adding planet formation depends mainly on the efficiency of planet formation itself, as well as the star formation history.

\begin{figure*}
\includegraphics[width=0.32\textwidth, clip=true, trim=4mm 4mm 4mm 3.5mm]{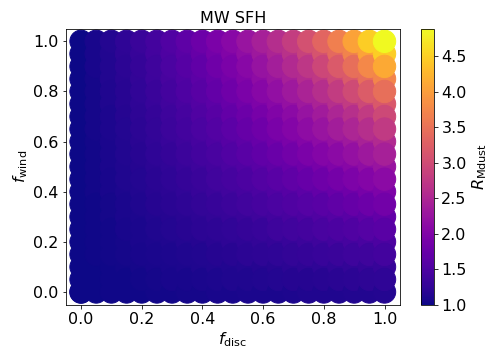}
\includegraphics[width=0.32\textwidth, clip=true, trim=4mm 4mm 4mm 3.5mm]{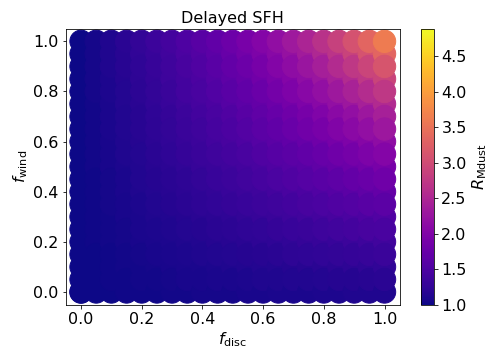}
\includegraphics[width=0.32\textwidth, clip=true, trim=4mm 4mm 4mm 3.5mm]{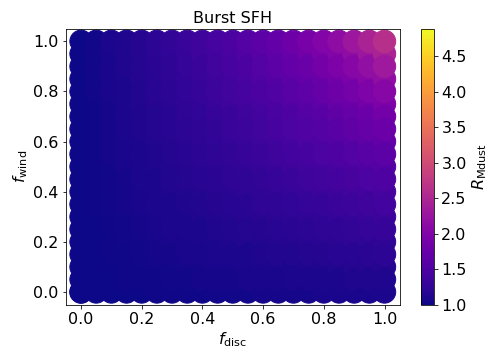}
\caption{The excess dust produced by planet formation compared to no planet formation (i.e. normal astration), for different values of $f_\mathrm{disc}$ and $f_\mathrm{wind}$. $R_\mathrm{dust}$ is calculated at time of maximum dust mass in the planet formation case (i.e. M$_\mathrm{dust},p = $max, the \emph{end of the SFH} in Figure~\ref{fig:Extradust_planets_time}). Three different star-formation histories are shown: Milky Way (left), delayed (middle), burst (right). The models assume no dust destruction by supernova shocks.}
\label{fig:Extradust_planets_max}
\end{figure*}

In Figure~\ref{fig:Extradust_planets_max} we show  $(f_\mathrm{disc},f_\mathrm{wind})$ as a function of $R_\mathrm{dust}$ at the end of the SFH, where M$_\mathrm{dust,p}$ typically reaches its maximum. In the Milky Way SFH, values of $f_\mathrm{disc}>0.7$ and $f_\mathrm{wind}>0.7$ result in more than a factor of two increase in the dust mass, and a factor of 4.5 increase for maximal planet formation. For the delayed and burst SFHs, slightly higher values of $f_\mathrm{disc}$ and $f_\mathrm{wind}$ are required to make more than a factor of two difference to the dust mass. At later times, we see that planet formation results in a factor of 2--4 increase in the dust mass, which may enough to make a significant difference to the dust budget, but only towards the later stages of each model galaxy's evolution.

The time evolution of the dust mass changes significantly if planet formation is active (Figure~\ref{fig:Extradust_planets_time}). In all cases where planet formation is inactive, the dust mass reaches a maximum before the present day, with the subsequent decline being due to astration by continuing star formation. If planet formation is active, then astration is reduced, and stars continue to produce dust by their associated planet formation.  This results in a continuous increase in $\mathrm{M_{dust}}$ as the present day is approached.

In summary, accounting for planet formation processes in chemical evolution models can only lessen the dust budget crisis in galaxies if most of the dust is not locked up in stars, the planet formation efficiency (i.e. the fraction of dust mass that ends up in planetary bodies) is relatively low and on timescales greater than a few Gyrs.

\subsection{Effects on the dust budget crisis at high redshift}

\begin{figure}
\includegraphics[width=0.48\textwidth, clip=true, trim=4mm 4mm 4mm 3.5mm]{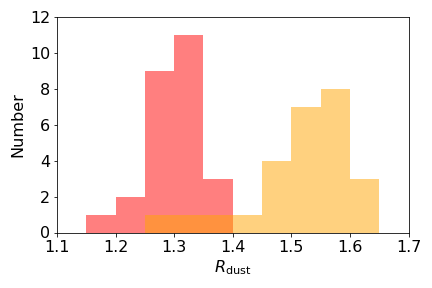}
\caption{The difference between the dust masses derived from the chemical evolution models with planet formation and no planet formation for the high-redshift SMGs, assuming dust is not destroyed by SN shocks. The red histogram is for $f_\mathrm{disc}=f_\mathrm{wind}=0.8$ and the orange histogram is for $f_\mathrm{disc}=f_\mathrm{wind}=1.0$. $R_\mathrm{dust}$ is calculated at time of maximum dust mass in the planet formation case (i.e. M$_\mathrm{dust},p = $max, usually at the end of each SFH).}
\label{fig:Extradust_planets_SMGs_hist}
\end{figure}

Having established that planet formation processes are only effective at boosting dust masses for high values of $f_\mathrm{disc}$ and $f_\mathrm{wind}$, we now consider the evolution of dust mass for 26 high redshift SMGs from \citet{Rowlands2014b} with $f_\mathrm{disc}=f_\mathrm{wind}=0.8$ and $f_\mathrm{disc}=f_\mathrm{wind}=1.0$. We compute the distribution of $\Delta \mathrm{M_{dust}}$ as the log difference in the observed and model dust masses of the SMGs at the end of the SFH, following \citet{Rowlands2014b}. $R_\mathrm{dust}$ is calculated at the end of the SFH in the same manner as the previous section (Figure~\ref{fig:Extradust_planets_SMGs_hist}). 

We find that if astration is reduced by 80\%, with strong disc winds ($f_\mathrm{disc}=f_\mathrm{wind}=0.8$) then on average the dust mass in the SMGs increases by a factor of 1.3. If we assume the maximal case where no metals and dust are locked into stars during the star and planet formation process (which is unphysical), then on average there is a factor of 1.5 increase in the dust mass of SMGs, compared to the case with no planet formation.

\begin{figure}
\includegraphics[width=0.48\textwidth, clip=true, trim=4mm 4mm 4mm 1mm]{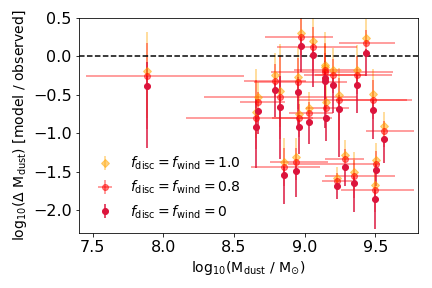}
\caption{The difference between the dust masses derived from the chemical evolution models with planet formation and the observed dust masses for the high-redshift SMGs, assuming dust is not destroyed by SN shocks. Dust is produced by low-intermediate mass stars and supernovae with no planet formation (dark red points), low-intermediate mass stars and supernovae with $f_\mathrm{disc}=f_\mathrm{wind}=0.8$ (light red points), with $f_\mathrm{disc}=f_\mathrm{wind}=1.0$ (orange points). Uncertainties on the observed dust mass are only shown for the light red points for clarity.}
\label{fig:Extradust_planets_SMGs}
\end{figure}

In Figure~\ref{fig:Extradust_planets_SMGs} we show the discrepancy between the dust masses derived from the chemical evolution models with and without planet formation, and the observed dust masses for the high-redshift SMGs. We assume that dust is also produced by low-intermediate mass stars and supernovae and that there is no dust destruction (i.e. a model with maximum stardust injection). 
As shown in \citet{Rowlands2014b}, with no planet formation (i.e. normal astration; dark red points), the model dust masses for the majority of the SMGs falls short of the observed dust masses. The model dust masses match the observed values (accounting for the $\sim\pm0.2$\,dex uncertainty in the observed dust masses) for 5/26 SMGs.

If we include efficient astration and disc winds ($f_\mathrm{disc}=f_\mathrm{wind}=0.8$, light red points), more dust is released back into the ISM as it is not locked up in stars. The increase in the dust mass is modest (a factor of 1.3 on average) and the model dust masses match the observed values for 8/26 SMGs. 

If we assume the maximal case where no metals and dust are locked into stars during the star and planet formation process (orange points) then the model dust masses match the observed values for 11/26 SMGs. On average the dust masses of the majority of the SMGs cannot be reproduced when dust is produced by low-intermediate mass stars and supernovae and when efficient planet formation reduces the amount of dust locked up in stars. This work adds to the growing body of evidence (e.g. \citealp{Morgan2003, Draine2009, Matsuura2009, Gall2011, Dunne2011, Boyer2012, Asano2013, Mattsson2014, Rowlands2014b, Zhukovska2014, Schneider2014, Wang17, DeVis2017} that dust sources such as interstellar grain growth via accretion are needed.

\section{Conclusions}
\label{sec:conclusions}

We have explored the possibility that the ``dust budget crisis'' in high redshift galaxies - the under-production of dust in galaxy chemical evolution models compared to observations - is due to these models neglecting the physical processes associated with planet formation which could reduce the amount of dust removed from the ISM when stars form (astration).

This paper displays the results of a galactic chemical evolution model updated to include a model of dust production via planet formation. This model introduces separate protoplanetary disc and protostar phases, and accounts for the reduced astration of dust and metals during star formation. This is because some material enters a protoplanetary disc, and is either a) ejected by jets, winds and outflows during the disc's lifetime, or b) assembled into kilometre-sized debris that can grind down to form dust on Gyr timescales, or c) assembled into planetary bodies that are no longer visible to high redshift observations.

We find that for the planet formation process to solve the dust budget crisis, it must be highly efficient at preventing astration, and most of the solid mass must be ejected from the system to account for the highly dusty nature of distant galaxies.  If planet formation efficiently hides solid mass in planet-sized objects, it can exacerbate the dust budget crisis. In the Solar System, the amount locked up in solids is a tiny fraction of a solar mass.  Also the power law distribution of grain sizes (Eq.~\ref{eq:size}) preferentially produces dust over planets, with natural processes eventually grinding the $\sim$ hundred km-sized material back to smaller grain sizes. These both imply that planets are a negligible part of the dust reservoir.

Planet formation (and associated protoplanetary disc processes) can only cause substantial dust mass enhancement at low redshift if star formation begins sufficiently early.  Increases in the dust mass of more than a factor of two is possible if the associated processes involved in planet formation are efficient at reducing the astration term, i.e. if the majority of solids entering a protostellar system enter a disc and are not accreted ($f_\mathrm{disc}>0.7$) and much of this dust is ejected from the system through jets and winds ($f_\mathrm{wind}>0.7$). This can solve the dust budget crisis using fiducial SFHs representative of low redshift galaxies. In high redshift SMGs, including planet formation in our chemical evolution models does not solve the dust budget crisis. Equally, our understanding of dust masses at $z=0$ places strong constraints on how efficient dust production due to planet formation processes can be, and suggests that this is unlikely to close the gap between models and observations of galaxy dust masses.

That being said, we admit our model of planet formation is necessarily simple, and does not reflect the full complexity of planet formation theory.  Our understanding of how protostellar discs of a variety of initial masses, radii and composition evolve into planetary systems is quite incomplete.  For example, future work may indicate a planet formation efficiency that varies significantly with redshift.  

We therefore recommend that future galaxy chemical evolution models continue to consider planet formation as a dust source, utilising data from the latest exoplanet and debris disc observations, as well as the best available theoretical models, to characterise the role of galactic ``microphysics'' on macroscopic scales.

\section*{Acknowledgements}

We thank the referee for their insightful comments on this work which have greatly improved the model. DHF gratefully acknowledges support from the ECOGAL project, grant agreement 291227, funded by the European Research Council under ERC-2011-ADG.  K.~R. acknowledges support from the European Research Council Starting Grant SEDmorph (P.I. V.~Wild).  HLG, LD and SM acknowledges support from the European Research Council (ERC) in the form of Consolidator Grant {\sc CosmicDust} (ERC-2014-CoG-647939, PI H\,L\,Gomez). LD and SJM acknowledge support from European Research Council Advanced Investigator Grant COSMICISM, 321302.  This research  has  made  use  of  NASA's  Astrophysics  Data  System Bibliographic  Services; {\sc GitHub} \url{https://github.com/}; Astropy 21 (a community-developed core Python package for Astronomy \citealp{Astropy2013}) and NumPy 24 \citep{Walt2011}.




\bibliographystyle{mnras} 
\bibliography{dustbudget_planets}







\bsp	
\label{lastpage}
\end{document}